\documentclass[aps,prl,floatfix,twocolumn,superscriptaddress]{revtex4-1}
\usepackage{amsmath,amssymb}
\usepackage{graphicx}
\usepackage{epsfig}
\usepackage{psfrag}
\usepackage[usenames]{color}
\usepackage{ulem}
\usepackage{soul}
\usepackage[bottom]{footmisc}

\begin{document}

%\preprint{}

%\title{Amplification of radiation in broad microwave and terahertz domain in graphene--superconductor hybrids}
\title{Optical Transistor for an Amplification of Radiation in a Broadband THz Domain}
% \\in graphene--superconductor hybrids}
%\title{Amplification of terahertz frequencies in graphene--superconductor hybrids}

%\title{Terahertz transistor based on graphene-superconductor hybrid}

%\title{Super-cooling of materials with light in graphene-superconductor hybrids}

%Optical response of hybrid graphene-superconductor system or Hybrid graphene-superconductor  for super-cooling of materials with light}

%\affiliation{Micro/Nano Fabrication Laboratory Microsystem and THz Research Center, Chengdu, Sichuan, China}

\author{K.~H.~A.~Villegas}
%\email[]{khvillegas@gmail.com}
\affiliation{Center for Theoretical Physics of Complex Systems, Institute for Basic Science (IBS), Daejeon 34126, Korea}

\author{F.~V.~Kusmartsev}\email[Corresponding author: ]{F.Kusmartsev@lboro.ac.uk}
\affiliation{Micro/Nano Fabrication Laboratory Microsystem and THz Research Center, Chengdu, Sichuan, China}

%\affiliation{ITMO University, St. Petersburg 197101, Russia}
\affiliation{Loughborough University,UK}
%\cortext{Corresponding author}
%\affiliation{*Corresponding author}

\author{Y. Luo}%\email[Corresponding author: ]{F.Kusmartsev@lboro.ac.uk}
\affiliation{Micro/Nano Fabrication Laboratory Microsystem and THz Research Center, Chengdu, Sichuan, China}

\author{I.~G.~Savenko}
\affiliation{Center for Theoretical Physics of Complex Systems, Institute for Basic Science (IBS), Daejeon 34126, Korea}
\affiliation{A. V. Rzhanov Institute of Semiconductor Physics, Siberian Branch of Russian Academy of Sciences, Novosibirsk 630090, Russia}
%\affiliation{Basic Science Program, Korea University of Science and Technology (UST), Daejeon 34113, Korea}

%\affiliation{*Corresponding author}


\date{\today}

\begin{abstract}
We propose a new type of optical transistor for a broadband amplification of THz radiation. 
It is made of a graphene--superconductor hybrid, where electrons and Cooper pairs couple by Coulomb forces. 
The transistor operates via the propagation of surface plasmons in both layers, and the origin of amplification is the quantum capacitance of graphene. 
It leads to THz waves amplification, the negative power absorption, and as a result, the system yields positive gain, and the hybrid acts like an optical transistor, operating with the terahertz light.   
It can, in principle, amplify even a whole spectrum of chaotic signals (or noise), that is required for numerous biological applications. % due to the wide range of target frequencies. 
 %\textcolor{red}{We believe the concept of optical transistor such as made of the graphene--superconductor  hybrid  can find multiple applications, in particular, in biological research and non-invasive diagnostics.}
\end{abstract}

\maketitle

%------------------------------------------
%------------------------------------------
%------------------------------------------

%\textit{Introduction.---} 
%\section{I. Introduction} 
%There has been a recent 
The  growing interest in terahertz (THz) frequency range (0.3 to 30 THz) is due to its potential applications in diverse fields
%, e.g. %. In particular, it can be used 
%in 
such as non-destructive probing in medicine, allowing for non-invasive tumor detection, biosecurity, %. Other applications include 
ultra-high bandwidth wireless communication networks, vehicle control, atmospheric pollution monitoring, inter-satellite communication, and spectroscopy ~\cite{Raab2002, Chamberlain2004, Eisele2010, Mittleman2018}.
However, %in spite of its technological potential, the 
the THz range still remains a challenge for modern technology due to the lack of a compact, powerful, and scalable solid state source~\cite{RefDragoman}. This problem is known as the \textit{terahertz gap}. 

%There have been multiple attempts 
To `close' this gap %and thus cover the full THz range. Approaching 
from the lower frequencies, one can mention electronic devices with negative differential resistance (NDR). %The NDR is required to compensate the thermal losses, so that the device can operate in the oscillating regime. 
For instance, super-lattice electronic devices (SLED) generate higher harmonics by means of NDR~\cite{Kusmartsev2001} and can reach 0.5 THz gap, while the output power is less than 0.5 mW ~\cite{Esele2018}. 
%
%So far, the highest frequency achieved with this method is 0.5 THz, and the output power is less than 0.5 mW. 
%Another class of devices is the resonant tunneling diodes (RTDs)~\cite{Asada2016}.  
%However, 
The radiation power of resonant tunneling diodes (RTDs)~\cite{Asada2016} is less than 1 $\mu W$, and it further decreases by three orders of magnitude at room temperature. Also, RTDs suffer from their small electron transition times and parasitic capacitance, associated with the double-barrier structure.

%Approaching the THz gap from the low-frequency side, one can also 
The use of layered high-temperature superconductors (HTSC) with intrinsic Josephson junctions, such as BISCCO~\cite{Ozyuzer2007, Welp2013, Sun2018} can %. These devices 
produce radiation with Josephson oscillations generated by an applied bias voltage~\cite{Marat1995, Marat2000}. 
Here a tunable emission, from 1 to 11 THz, has been recently observed~\cite{Krasnov2017}. However, the power output is  1~$\mu$W, which is still inadequate for practical applications.

%There exist other proposals of THz devices. It has been recently shown that the 
%It was proposed that a spontaneous emission rate of THz radiation in a semiconductor microcavity 
It can be enhanced with the use of Bose-Einstein condensates~\cite{RefKavokin, RefPRLTHz}. However, such approach requires a hybridization  of several bands with different parity, making the output power small.
%
%Other coherent sources of THz radiation include 
Quantum cascade lasers (QCL)~\cite{Kohler-2002, Williams-NP-2007,Faist2013} %are devices that 
can generate a high-frequency THz-radiation, while transistors~\cite{Hubers-2005, Chas-2009, Lus-2005, Li-2010}, Gunn diodes~\cite{gunn-diodes}, and
frequency multipliers~\cite{freq-mult} are approaching the THz-gap from the low-frequency side. The latter covers the whole THz range while having small power.
The general fundamental obstacle of all these THz sources is the small emission rate (of the order of 10 ms). It can be increased with the Purcell effect when the THz source is placed in a cavity~\cite{RefTodorov, RefChassagneux}, however, the quantum efficiency is still about $1\%$, and the manufacture of these devices is difficult.

Graphene and carbon nanotubes may serve as %in 
%open a perspective to 
highly tunable sources and detectors of THz radiation~\cite{Portnoi2006, Portnoi2007, Portnoi2008,  Mikhailov2009, Portnoi2010,  Portnoi2014, Ryzhii2015}, and  %When exposed to light, graphene displays NDR, which can be used in 
even in THz lasers~\cite{Ryzhii2016,Ryzhii2011,Ryzhii2012,Ryzhii2013,Ryzhii2012A,Li2012}. 
In the dual-gate graphene-channel field-effect transistor~\cite{Ryzhii2011A} embedded into a cavity resonator~\cite {Ryzhii2018, Hramov2014}, one observes spontaneous broadband light emission in the 0.1-7.6 THz range with the maximum radiation power of $\sim10~ \mu$W at the temperature 100~K. %Still, a much higher power is required for numerous applications.
There are also emerging sources of THz-radiation that  can deliver several frequencies at room temperature, e.g., %and thus broadband, 
multiple harmonic generation in superlattices %that can also be made compact by coupling with a sled  
\cite{Pereira2017,Apostolakis2019}, frequency difference generation in mid infrared QCLs %leading to tunable THz radiation
~\cite{Razeghi2015} and  THz optical combs \cite{Forrer2018}.

%re \textcolor{blue}{(8)}The amazing properties of carbon nanomaterials such as graphene and carbon nanotubes, especially its THz plasmonics, open the door for potential innovation in optoelectronic devices\cite{Portnoi2006,Portnoi2007}. The anisotropic optical properties of carbon nanotubes can be used for THz antennas and polarizers\cite{Ren2012}. The ability to control their electronic properties via external electric and magnetic fields may be used  to build a highly tunable sources and detectors of THz radiation\cite{Portnoi2008,Mikhailov2009,Portnoi2010,Portnoi2014}. Furthermore, graphene's transport properties, which are based on massless electrons, allow highly sensitive detection and emission of THz radiation\cite{Ryzhii2015}. Under optical irradiation, graphene displays negative differential conductivity(NDC) which provides THz gain and may lead to new types of THz lasers\cite{Ryzhii2016,Ryzhii2011,Ryzhii2012,Ryzhii2013,Ryzhii2012A,Li2012}. 

%
%
%
\begin{figure}[t!]
\includegraphics[width=0.468\textwidth]{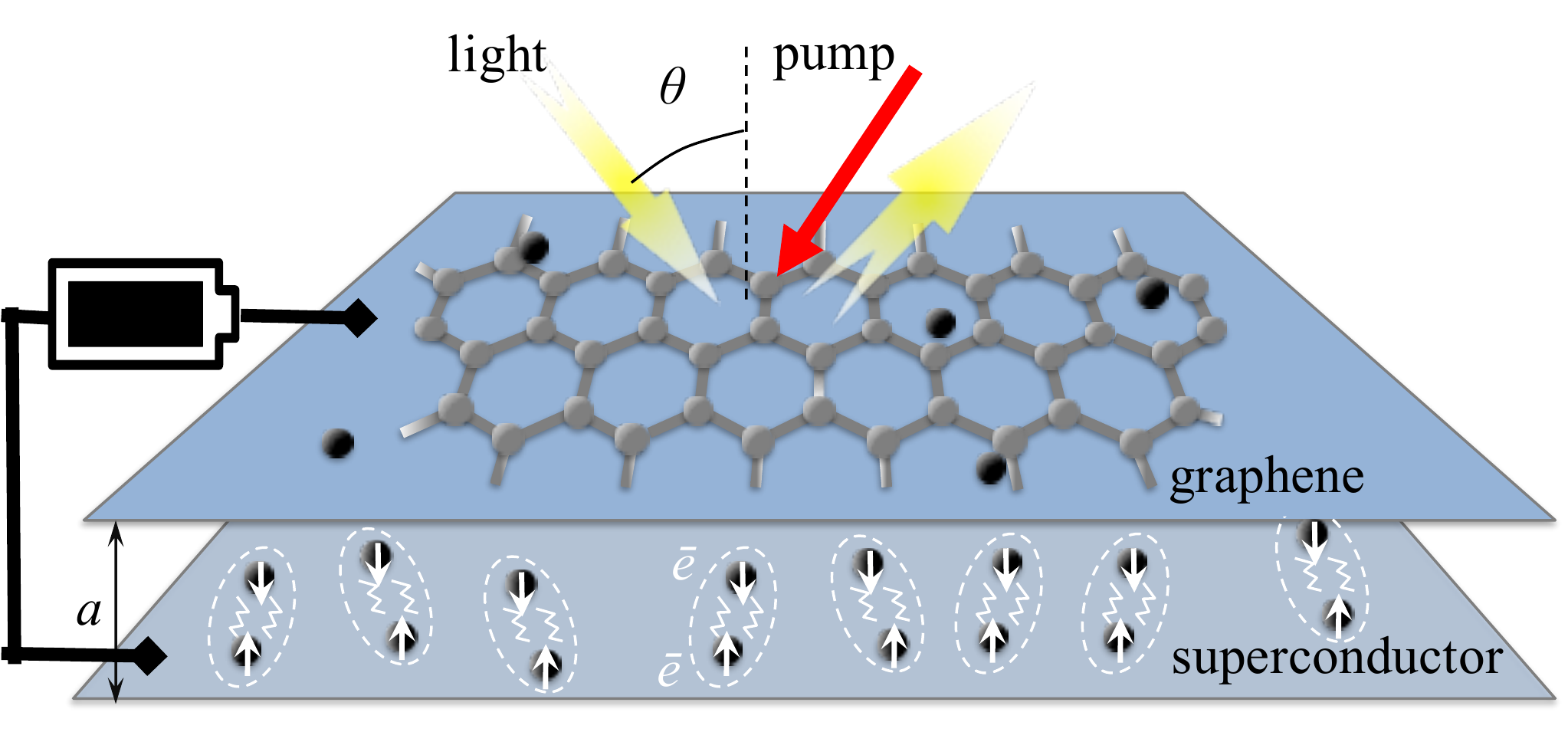}
% \includegraphics[width=0.49\textwidth]{operation1.pdf}
% \begin{figure*}[t!]
%\centering
%\includegraphics[width=9cm]{operation1.pdf}%[width=8.5cm]
\caption{\label{Fig1} 
System schematic. Graphene coupled with a two-dimensional superconductor by the Coulomb force and connected to an electrical pump source (a battery). %The hybrid system is exposed to an external EM field at incidence angle $\theta$; 
\textcolor{black}{Figure also shows the pump-probe configuration for the THz radiation amplification: The hybrid system is exposed to an external laser (\textit{pump}, depicted by red arrow) and broadband EM field at incidence angle $\theta$ (\textit{probe}, depicted by yellow arrows). 
The frequency of the pump (probe) should be above (below) the superconducting gap. 
Both the optical and electrical pump can provide energy for the amplification. }
}
\end{figure}
Graphene covered with a thin film of colloidal quantum dots %shows
has strong photoelectric effect, that provides enormous gain for the photodetection (about $10^8$ electrons per photon)~\cite{Konstantatos2012}; graphene grown on SiC has strong photoresponse~\cite{Trabelsi2017}; and graphene composites can improve solar cells efficiency~\cite{Bkakri2015}. 
%In this Letter, 
Note, both graphene and superconductor alone are practically insensitive to light~\cite{Yung2013}.
In this Letter, we show that graphene placed in the vicinity of  % in the vicinity of a 
a superconductor represents an active media with strong light-matter coupling. It  can operate as %and used for %. 
%Making 
an optical transistor that amplifies broadband electromagnetic radiation. 
%The optical transistor 
%and  deliver 
%a high power. %at low temperatures. and 
 %that 
%opening a way for numerous applications of THz radiation. 
%
%To compare the proposed 
%Such optical transistor %with the 
%As THz QCL %quantum cascade lasers, %we, first of all, acknowledge that QCLs 
%possesses a number of outstanding properties.% such as
%In particular, 
%they 
%it 
%They can be 
%used as amplifiers exploiting a relatively broad gain. However, as we will demonstrate, the optical transistor based on superconductor-graphene hybrid can 
%amplify %the whole broadband spectrum, including 
%a very broad frequency spectrum. % of noise. 
This new feature %is new and it 
%opens a possibility 
allows one to use this device for studying chemical and biological processes or in telecommunications for encryption-decryption procedures, where it is important to image the whole spectrum.

%------------------------------
%------------------------------
%------------------------------

%\section{II. Theory of light absorption}
%{\bf Theory of light absorption.} 
We consider a system consisting of parallel layers of graphene and a superconductor, exposed to an electromagnetic (EM) field incident with the angle $\theta$
and linearly polarized along the x-z plane (p-polarisation),
$\mathbf{E}(\mathbf{r},t)=\mathbf{%\hat{x}
(\sin\theta,0 ,\cos\theta)}E_0 e^{-i(k_{\perp}z+\mathbf{k}_\parallel\cdot\mathbf{r}+\omega t)}$,
where $\mathbf{k}_\parallel$, $\omega$, and $\mathbf{r}$ are the in-plane wave vector of the field, frequency, and coordinate, respectively (see Fig.~\ref{Fig1}). Between  graphene layer and superconductor there is a gate voltage that %\textcolor{red}{
controls its chemical potential \textcolor{black}{and provides AC power. It may pump the energy in the superconductor with AC bias, larger than the superconducting gap exciting quasiparticles and creating NDR.
Alternatively, the battery can be replaced by an external laser (pump) with frequency   exceeding the superconducting gap (Fig.~\ref{Fig1}).}

The electrons in graphene are coupled by the Coulomb interaction, which has the Fourier image given by $v_k=2\pi e^2/k$, where $\mathbf{k}$ is in-plane momentum (lying in the x-y plane). The electrons between the two layers are also Coulomb-coupled, and the Fourier image of the interlayer interaction reads $u_k=2\pi e^2\exp(-ak)/k$, where $a$ $\sim 10$ nm is the separation between the layers.

\begin{figure*}[t!]
\includegraphics[width=18cm]{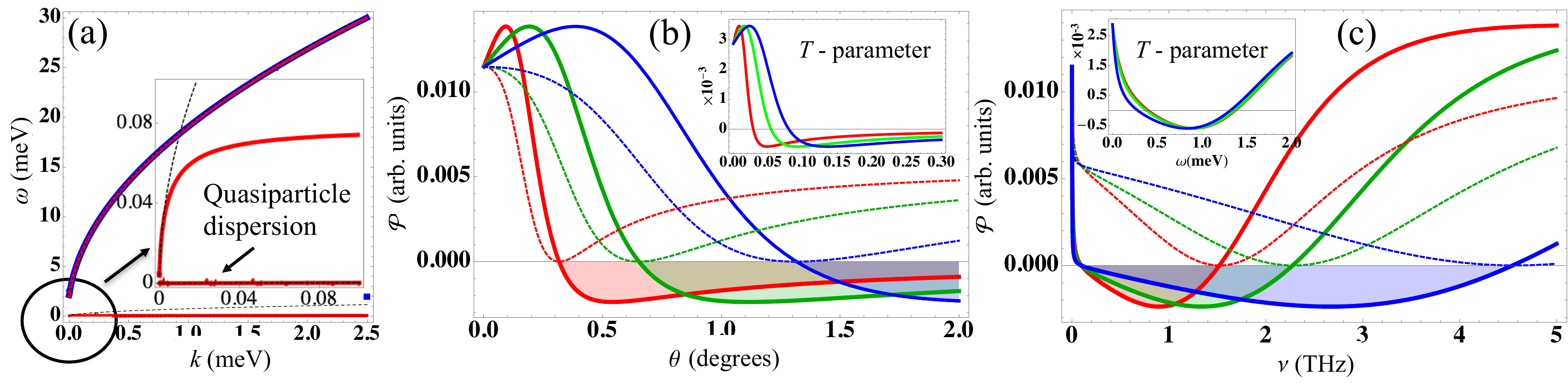}
\caption{\label{Fig2} (a) Hybrid collective plasmonic modes for undoped (blue curves) and doped (red curves, $\mu=3.0$ meV) graphene layer. Black dashed curves show the modes of isolated doped graphene layer. Inset. zoom-in of the lower energy modes. 
(b) Power absorption as a function of the angle of incidence $\theta$ [see also Fig.~\ref{Fig1}] for %undoped 
graphene-superconductor hybrid for frequencies $\nu=0.5$ THz (red), $1.0$ THz (green), and $2.0$ THz (blue). (c) %Undoped 
Graphene power absorption spectrum for the angles of incidence $\theta=1.0^\circ$ (red), $\theta=1.5^\circ$(green), and $\theta=3.0^\circ$(blue). Dashed curves show the data corresponding to the isolated graphene case. In (b) and (c) $\mu=0$; in insets, the effect of temperature $T=0$ (red), $T=0.5T_c$ (green), and $T=T_c$ (blue) is shown. The separation between graphene and superconductor is $a=10$~nm.
}
\end{figure*}

Using the linear response theory for hybrid systems~\cite{RefOurFano, RefOurParamag}, we can write the electron density fluctuations in the graphene layer $\delta n_{k\omega}$ and Cooper pair density fluctuations in the superconducting layer $\delta N_{k\omega}$ as~\cite{Villegas2018}
\begin{eqnarray}
\label{lr}
\delta n_{k\omega}&=&\Pi_{k\omega}(v_k\delta n_{k\omega}+u_k\delta N_{k\omega}+W_{k\omega}),\nonumber\\
\delta N_{k\omega}&=&P_{k\omega}(v_k\delta N_{k\omega}+u_k\delta n_{k\omega}+W_{k\omega}),
\end{eqnarray}
where $\Pi_{k\omega}=\Pi_{k\omega}^R+i\Pi_{k\omega}^I$ and $P_{k\omega}=P_{k\omega}^R+iP_{k\omega}^I$ are the complex-valued polarization operators of the graphene and superconductor, respectively, and $W_{k\omega}={eE_0}/{ik}$ is the Fourier image of the potential energy due to the external electric field. From Eqs.~(\ref{lr}) we can find the density fluctuations in graphene $n_{k\omega}$ and in superconductor $N_{k\omega}$ as linear functions of applied electric field amplitude $E_0$ (see Supplemental Material~\cite{SM} for details). 
Collective plasmonic hybrid modes in graphene and superconductor can be found from the same 
system of equations, taking into account the expressions for the polarization loops of the superconductor~\cite{Villegas2018} and graphene~\cite{Wunsch2006, Hwang2007}. Substituting the expressions for $n_{k\omega}$ and $N_{k\omega}$ into the continuity equations, $kj_{k\omega}=-e\omega\delta n_{k\omega}$  and $kJ_{k\omega}=-2e\omega\delta N_{k\omega}$, for graphene and superconductor, respectively, we can determine the electric currents in each of the layers and their impedances $Z_{G}$ and $Z_{SC}$.
%\section{III. Results and discussion}
%{\bf  Results and discussion}. Substituting Eqs.~\eqref{EqPolS} and~\eqref{EqPolG}  into Eq.~\eqref{det2}, we can find 
The collective modes of the hybrid system are presented in Fig.~\ref{Fig2}(a) for the undoped and doped graphene cases. %(for a detail, see the supplementary material). 
The upper mode has a gap $2\Delta=2$~meV. If in this hybrid a single graphene layer is not interacting with a superconductor, only one mode exists, which is due to the superconductor. 
%
%
%

%The next task is to find the polarization operators. 
%For the superconductor, the polarization is given by~\cite{Villegas2018}
%
%\begin{eqnarray}
%\label{EqPolS}
%\nonumber
%P^R_{k\omega}&=&\frac{p_{SF}^2k^2}{2\pi m_S}\cdot\frac{1}{\omega^2-4\Delta^2},\\
%P^I_{k\omega}&=&
%\begin{cases}
%0, & 0\leq\omega\leq 2\Delta\\
%-\frac{p_{SF}^2k^2}{2m_S\omega}\cdot\frac{1}{\sqrt{\omega^2-4\Delta^2}},& 2\Delta<\omega
%\end{cases}
%\end{eqnarray}
%
%where $p_{SF}$ is the Fermi momentum, $m_S$ is the effective electron mass in the superconductor, and 2$\Delta$ is the gap. 
%
%For graphene, the polarization reads~\cite{Wunsch2006,Hwang2007}
%
%\begin{eqnarray}
%\label{EqPolG}
%\Pi_{k\omega}&=&-\frac{i\pi}{v_F^2}F(k,\omega)-\frac{g\mu}{2\pi v_F^2}%\nonumber\\
%+\frac{F(k,\omega)}{v_F^2}\bigg\{G\left(\frac{\omega+2\mu}{v_Fk}\right)\nonumber\\
%&{}&{}-\Theta(y-1) \left[G(y)-i\pi\right]-\Theta(1-y) %\nonumber\\
 %G\left(-y\right)\bigg\},
%\end{eqnarray}
%
%where $G(x)=x\sqrt{x^2-1}-\mbox{ln}(x+\sqrt{x^2-1})$;  $\Theta(x)$ is the Heaviside step function, $y=(2\mu-\omega)/(v_Fk)$, and  %the function 
%
%\begin{eqnarray}
%F(k,\omega)=\frac{g}{16\pi}\frac{v_F^2k^2}{\sqrt{\omega^2-v_F^2k^2}},%\nonumber\\
%G(x)&=&x\sqrt{x^2-1}-\mbox{ln}(x+\sqrt{x^2-1}),
%\end{eqnarray}
%
%where $v_F$ is the Fermi velocity in graphene, $\mu$ is the chemical potential, and $g=4$ is the spin and valley degeneracy factor. 
%For the undoped case, $\mu=0$ and only the first term in Eq.~(\ref{EqPolG}) contributes.

%%%%%%%%%%%% cut cut cut %%%%%%%%%%%%%%%%%%%%%%%%%

The formula for the power absorption or gain reads~\cite{Landau}
\begin{eqnarray}
P(\omega)=\frac{1}{2}\left\langle\mathcal{R}e\left[\int d^2r\mathbf{J}(\mathbf{r},t)\cdot\mathbf{E}^*(\mathbf{r},t)\right]\right\rangle,
\end{eqnarray}
where the integration is over the graphene plane, $\langle\cdot\cdot\cdot\rangle$ denotes time-averaging. %, and the current $\mathbf{J}$ is monitored in the graphene layer. 
We %use the continuity equation $kj_{k\omega}=-e\omega\delta n_{k\omega}$ and
 normalize the power with the sample area $\int d^2=l^2$ and the square of the field amplitude $E_0^2$ to get
\begin{eqnarray}
\label{NormPower}
\mathcal{P}(\omega)=\frac{P(\omega)}{l^2E_0^2}=\frac{1}{2}\frac{e\omega}{kE_0}\mathcal{R}e\big[\delta n_{k\omega}\big].
\end{eqnarray}

%---------------------
%---------------------
%---------------------

%Furthermore we build the power absorption~\eqref{NormPower}, fixing $\mu=0$ at the Dirac point by gate voltage.  %focusing on the undoped graphene case only
%. In this case, the features of
%Graphene is ambipolar gapless semiconductor with linear dispersion and zero-area Fermi points.  % -- are emphasized.
%
Figure~\ref{Fig2}(b) shows the dependence of the power absorption %function 
on the EM field incidence angle $\theta$ calculated  with~\eqref{NormPower}, fixing $\mu=0$ at the Dirac point by gate voltage. All the curves exhibit critical angles at which the power absorption becomes negative, $\alpha<0$. This suggests that the incident angle can be used to switch the amplifier device on or off. Furthermore, increasing the frequency of the incident EM wave increases the critical angle. 
% It is positive for all the angles when the incident field energy is less than the superconducting gap $\omega<2\Delta=2$ meV (red and green curves). For $\omega=3$ meV, the power absorption of the hybrid system becomes negative at the critical angle $\theta_c=0.26^\circ$ (see blue curve and the shaded region). That is in a strong contrast with the isolated free standing graphene (dashed blue curve), where it stays positive for all the angles. When the light incidence angle is larger than  $2^\circ$, all the light  absorption curves  saturate fast approaching to some constant values as $\theta\rightarrow 90^\circ$ and therefore we only show the interval $0$--$2^\circ$. 

Figure~\ref{Fig2}(c) shows the power absorption spectrum. We see that coupling graphene to the superconductor layer results in a negative power absorption in THz frequency range (solid curves and the shaded regions). 
There is no negative absorption region for isolated graphene, where the power absorption remains positive for any frequency $\nu$ (dashed curves).
%
%With the increase of the angle of 
When the light incidence angle $\theta$ increases, both the maximal intensity (slightly) and the frequency range  of the negative light absorption increase, see the shaded area in Fig.~\ref{Fig2}(b,c). 
Thus, the angle of light incidence allows us to control the range of light frequencies with the negative absorption.

To understand the $\theta$-dependence, note that the wave vector of the plasma wave is related to the projection of the incident light wave vector on the plane of the sample. 
Both the angular dependence of the absorption and gain are related to the amplitude of this wave propagating on the surface. 
The light incident perpendicular to the graphene surface can not excite such plasma waves and therefore, in this case we do not have the gain. 
However, at large incident angle, there is a reflection of the incident radiation due to the difference in the refractive index of the hybrid and air. 
Thus we conclude that the most optimal effect will be observed at small but nonzero $\theta$.
%The gate voltage attached to the graphene layer acts both as electron sink and source. %Due to
 % graphene layer tends to lower its energy level by dumping the electrons to this sink resulting in a lower Fermi level, with the energy carried away by the emitted radiation. The same gate voltage then also acts as a source to replenish the electrons bringing the Fermi level back to the Dirac point. 
If the system is embedded into a cavity resonator, there might even arise lasing similar to one observed in plasmonic lattices~\cite{remark-X, Zhou2013} or semiconductor superlattices~\cite{Hramov2014}.

%------------------------------
%------------------------------
%------------------------------

%\section{IV. Discussion}
%Plasmons can be treated as bosons obeying the Bose-Einstein distribution. %~\cite{Einstein1925}. 
%However under external irradiation there can appear a virtual inversion of their population and their continuous stimulated emission (positive gain) accompanied by the amplification of the incident light. In our case, it corresponds to $\alpha<0$. 
%
%The resonant frequency $\omega_0$ separates the positive gain region from the negative one. When the photon frequency $\omega$ is below the resonance, the system gives energy away, and the gain is positive, while if $\omega$ is larger than the plasmon resonance frequency, the light absorption is positive and there is no gain. At special conditions e.g., if the system is embedded into a cavity resonator, there may even arise lasing similar to one observed in plasmonic lattices~\cite{Zhou2013}.

%The obtained results can be phenomenologically explained in a similar way as it was done for other hybrid system -- 
The mechanism of gain here is similar to one in a waveguide coupled with a superconducting Josephson junction~\cite{Pedersen2009}. %the gain
 %Here we deal with graphene instead of a waveguide and a superconducting film instead of a Josephson junction.
 %If we denote the impedance of graphene as $Z_1=R_1+ i X_1$ and the impedance of the superconducting film as $Z_2=R_2+ i X_2$, 
Then, the optical reflectivity of the system reads $\Gamma=(Z_G-Z_{SC})^2/(Z_G+Z_{SC})^2$. Near the frequency of the plasmon resonance, there is an area of negative differential resistance of the superconductor, $R_{SC}<0$. 
If we assume $X_G=0$ and $X_{SC}=0$~\cite{Pedersen2009,remark10}, we find $\Gamma>1$. Note, that graphene can also have NDR (see Sec.~III of~\cite{SM} and~\cite{Kummel1990,Hyart2008,Hyart2009}) as
%
%Note that 
%The similar physics exists 
in a graphene transistor, which  consists of two %of a couple of 
graphene layers separated by a BN insulator~\cite{Britnell2013}.  

\textcolor{black}{The graphene-superconductor junction (Fig.~\ref{Fig1}) has large tunneling resistance. An electron in graphene with energy below the superconducting gap can tunnel into superconductor only due to the Andreev scattering \cite{Andreev1964}. The probability of such tunnelling is small since all electrons are paired. Therefore, the resistance of the junction is high.  
With applied bias voltage above the gap, quasiparticles appear and they can  tunnel. As a result, the resistance decreases and NDR arises. The latter can appear even at zero bias when we pump the superconductor with external light with the frequency above the gap.  
The light  excites electron and hole quasiparticles coexisting with superconducting fluctuations on the surface of the superconductor~\cite{Aslamazov1968}. Then,  in addition to the Andreev 
 scattering, there starts normal tunneling of quasiparticles into the superconductor.  The resistance of the junction decreases and the NDR arises. Such mechanism of NDR can exist only in a highly nonequilibrium excited state created by the pump.} 

Scattering and de-phasing mechanisms are limiting the gain bandwidth and can flat and eliminate the gain %see, for a detail, Ref. 
\cite{Pereira}. %The qualitative discussion of the de-phasing and scattering mechanisms is given in Ref. \cite{Pereira}.
%In our system 
Here the plasmon scattering within each and between two graphene and superconducting layers can not only broaden the width of the optical transition, but also enable optical gain and absorption to coexist, constituting the Wacker-Pereira mechanism of optical gain~\cite{Wacker2007,Wacker2012,Pereira}, as observed in QCLs%quantum cascade lasers
~\cite{Terazzi2007, Revin2008}. 
It provides one of the explanations, why $\alpha(\omega)=1-\Gamma(\omega)$ is negative below the plasmon resonance. 

%
%
%
%
%
%
%
%

%We can also look at all this from a slightly different perspective. Let 
From another perspective, %the superconductor be Nb, Pb, or some high-temperature superconductor. The two layers are separated by a dielectric layer made of SiO$_2$ or Ta$_2$O$_5$. 
%The 
graphene separated by dielectric layer (e.g., made of BN, SiO$_2$ or Ta$_2$O$_5$) from the superconductor (Nb, Pb, or HTSC) together form a parallel plate capacitor, which 
%\textcolor{red}{%The total 
capacitance $C$ %of this graphene-superconductor hybrid 
is given by
${C}={C_{plate}}+{C_{q}}$,
%${1}/{C}={1}/{C_{plate}}+{1}/{C_{q}}$
where $C_{plate}$ is the classical capacitance $C_{plate}=\epsilon_0 A/a$, %where 
$A$ is the area of the sample and $\epsilon_o$ is the dielectric constant (e.g., $\epsilon_o=3.9$ for SiO$_2$). 
The quantum capacitance $C_{q}$ of graphene emerges due to its conical energy-momentum relation, and it has the form
%
%\begin{equation}
$ C_{q} =  %\frac
{2 A e^2  |E_F |}/{\pi \hbar^2 v^2_{F}}$~\cite{Yu, Trabelsi2014, Trabelsi2016}.
%\end{equation}
%

\begin{figure}[t!]
\includegraphics[width=0.468\textwidth]{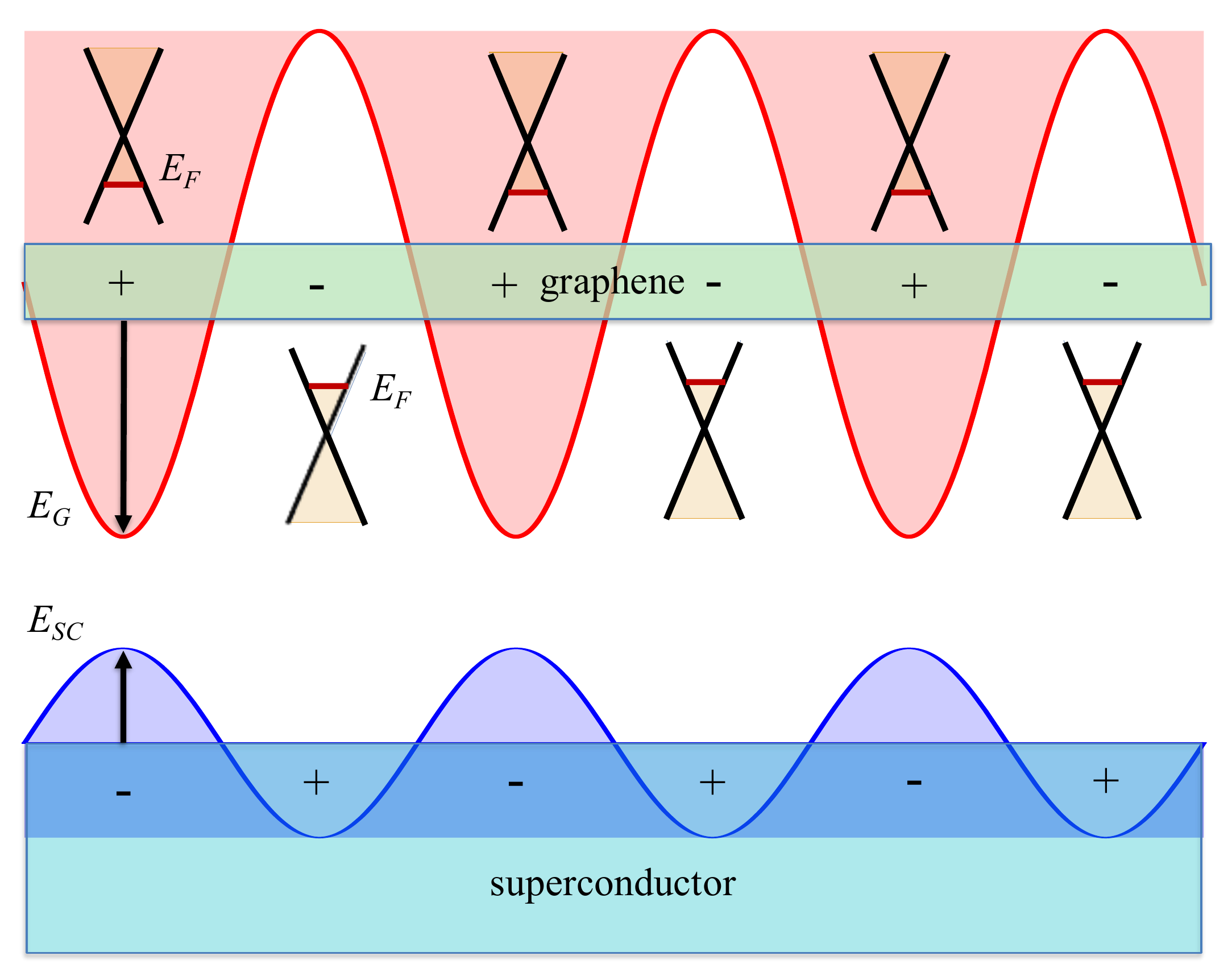}
% \includegraphics[width=0.49\textwidth]{operation1.pdf}
% \begin{figure*}[t!]
%\centering
%\includegraphics[width=9cm]{operation1.pdf}%[width=8.5cm]
\caption{\label{Fig3} 
Schematic of the mechanism of THz amplification. The incident light induces a collective hybrid plasmon mode. The interplay between this mode and the quantum capacitance of graphene amplifies the incident electromagnetic field (see text for details).
}
\end{figure}

This parallel plate capacitor is connected to a power supply that charges the capacitor and provides the energy for the amplification of incident radiation.  The incident light (in particular, its component parallel to the superconducting surface) induces the fluctuation of charge density %on superconducting surface 
$ \delta n_s$, which is associated with a travelling plasmon wave with the  amplitude $E_s\sim  \delta n_s$, where $\delta n_s=\delta n_{s0} \cos({kx+\omega t})$ (Fig.~\ref{Fig3}). 
This  charge density wave on the surface of the superconductor generates a mirror charge wave of the opposite sign in the neutral graphene layer, being of the same order as the charge fluctuations in superconducting layer, i.e. $\delta n_G\sim \delta n_s$.  Note that  the sign of these charge density fluctuations changes each half-wavelength of the plasmon wave. These plasmons have the wavelength larger than a micrometer, and therefore the charge density for each half-wavelength can be viewed at as a local temporary graphene doping, moving with the wave. During this half-period, the charge fluctuation corresponds to the local change in the chemical potential or the Fermi energy $E_F$,  which is directly related to the amplitude of the plasmon wave propagating in graphene $E_G$.  Due to the quantum capacitance of graphene, described by the relation $ E_F \sim   \sqrt{n_G}$, the waves amplitude,  $ E_G\sim E_F$, is significantly 
enhanced and it is different from the amplitude of the plasmon wave, propagating on the surface of the superconductor.
 %There occurs a huge enhancement of the wave amplitude already, when it is propagating in  graphene.  
%The huge enhancement %of the waves amplitude 
%of this wave amplitude is related to the graphene %charge density, which is due to its 
%quantum capacitance described by the relation $ E_G\sim E_F/\lambda\sim  \sqrt{n_G}$. 
%This is illustrated in Fig.~\ref{Fig3}, where the plasmon wave on the surface of superconductor, which amplitude $\sim \delta n_s$ is shown in blue,  while its plasmon counterpart in graphene enhanced due to the quantum capacitance with the amplitude $\sim \sqrt{ |\delta n_s|}$  is shown in red.
% 

%
%
%

\textcolor{black}{In order to stimulate radiation of electromagnetic waves in the THz range (similar to photo-conductive antennas~\cite{Burford2017}), one has to add  a supply of energy. It can be done by introducing a  pump-probe setup, where the  energy required for  the amplification of the probe comes from the pump.  For the pump, we have to  expose graphene to an external laser with the frequency above the superconducting gap. Or we can apply AC bias voltage with the amplitude larger than the gap. 
Due to pump, there forms a state, in which energy is accumulated in electronic excitations. To have the pump-probe configuration, we expose the junction to an additional external light (probe) with the frequency lower than the gap. Then, there on the surface will emerge superconducting charge density fluctuations. Effectively they represent coherent waves traveling inside the capacitor formed by graphene and  superconductor. Due to the graphene quantum capacitance~\cite{Yu, Trabelsi2014, Trabelsi2016}, the amplitude of these electromagnetic waves becomes amplified (by pump), resulting in reflection coefficient larger than 1 (as shown above). 
%The energy for this amplification is taken from the charge fluctuations created by the external  pump.
% Therefore, there is no breaking of any laws of thermodynamics in this process. The critical temperature $T_c$ and the value of superconducting gap, $\Delta$, limit  working temperatures and  the gain bandwidth, respectively, for the proposed optical transistor.
} 

Note that the graphene--superconductor system possesses many benefits.
Both superconducting and graphene layers have giant mobility and small resistivity giving rise to minute losses. 
%Therefore, the hybrid can be an extremely efficient source of radiation.    
With increasing temperature the  superconducting gap decreases, while graphene mobility changes a little. Since the radiation amplification occurs in graphene the temperature has a weak effect on the operation of THz transistor (see insets in Fig.~\ref{Fig2}(b,c). 
 %Since the superconducting gap $\Delta$ decreases with the temperature and the penetration depth $\lambda$ increases when the temperature rises, the optimal operation of the proposed device will be somewhere around the middle of critical superconducting temperature, ie ~ $T_c/2$ .  Therefore, 
The  working temperature range is similar to one %the temperature range of 
for the  stack of the Josephson junctions made of HTSC~\cite{Ozyuzer2007,Welp2013} and %However, obviously, the operation temperature range
it is limited by the critical temperature $T_c$.  %On the other hand, the optimal range of quantum cascade lasers in THz range is limited  by the energy of optical phonons and it is of the same order. However, QCL is not operating  for sub-THz frequencies. Therefore, the range of working temperatures and frequencies are complementary to ones associated with quantum cascade lasers. (see, e.g. \cite{Williams-NP-2007,Faist2013}).

The radiation power reads $P_r=\langle V_g I\rangle$, where the voltage $V_g$ and the transverse current $I$ are periodically changing in time around the Dirac point. Then, assuming simple periodic behavior for both $I(t)$ and $V_g(t)$ and  the graphene-superconductor separation, $a$=10 nm \cite{remark-distance}, the maximal outcome power reads $\langle I \times V_g\rangle\sim$~200-250~$\mu$W/cm$^2$~\cite{remark-power}. Evidently, it can be increased for larger areas of the surface or employing multilayer hybrid structures.

{\bf Conclusions.} We have shown that in a hybrid graphene--superconductor system exposed to an electromagnetic field of light the absorption coefficient can become negative in a certain range of frequencies and at a non-zero angle of incidence. 
We suggest that the system can serve as an amplifier of THz radiation. The essence of the amplification is the quantum capacitance of graphene, which %serves for
provides the conversion of the charge density wave induced by incident light into emitted radiation with much stronger intensity. That is also
%This effect also 
related to the negative differential conductivity of the hybrid, where there is a strong Coulomb coupling of graphene and superconductor. 

Such devices are now in strong demand and may be complementary to quantum cascade lasers. 
Moreover, the usage of high-temperature superconductors extends the range of temperatures required for their operation. 
%of such hybrid devices and that  can be achieved at liquid nitrogen.
%-------------------
%-------------------
%-------------------
%\section{Outlook} 
%{\bf Outlook.} 
The existence of Dirac or Weyl cones in graphene, topological insulators, and Weyl semimetals brings in a new physical concept called quantum capacitance. Its essence is in a strong dependence of the Fermi energy on the charge doping. A weak charge density wave can induce a strong electric field in these materials, allowing us to achieve the amplification of incident electromagnetic radiation. 

The situation is somewhat similar to lasers, where the pumping results in the population inversion. The difference is that here the amplification can occur in a broad frequency range simultaneously,  while in lasers it is pinned to a specific resonant frequency. Such amplification of the broadband spectrum, e.g., for chaotic or noise radiation, opens exciting opportunities of new types of molecular and biological noise spectroscopy,  where the response of the system can be measured in a broad frequency range opening new opportunities in molecular and biological noise spectroscopy~\cite{Koshelets-IEEE-2015,Gulevich-2017}.

%------------------
%------------------
%------------------

%\section{Acknowledgements} 
\textit{Acknowledgements.} We thank Vadim Kovalev and Gennadii Sergienko for fruitful discussions. K.H.V. and I.G.S. acknowledge the support of the Institute for Basic Science in Korea (Project No.~IBS-R024-D1) and the grant RFBR~18-29-20033~mk.
%V.~M.~K. has been supported by the Russian Foundation for Basic Reaserch (Project No.~16-02-00565).
The work of F.V.K. was supported by the Government of the Russian Federation through the ITMO Fellowship and Professorship Program.

%
%
%
%- - - - - - - - - - - - - - - - - - - - - - - - - - - -- -- -- - - - - - - - - -- - - - - - - - -
%- - - - - - - - - - - - - - - - - - - - - - - - - - - -- -- -- - - - - - - - - -- - - - - - - - -


\begin{thebibliography}{9}
\bibitem{remark-X} Plasmons can be treated as bosons obeying the Bose-Einstein distribution. %~\cite{Einstein1925}. 
However under external irradiation there can appear a virtual inversion of their population and their continuous stimulated emission (positive gain) accompanied by the amplification of the incident light. In our case, it corresponds to $\alpha<0$.
\bibitem{remark10} Note, that here the absorption and gain have been calculated  directly as the power of the absorbed or emitted electromagnetic radiation with Eq. (8),  on the basis of the solutions of  Eqs. (1-6) . The concept of impedance has not been used in this calculation. However,  from the same equations, one can find impedances of graphene and SC as coefficients proportional to density fluctuations of graphene and superconductor, or  AC electric currents expressed through these density variations, respectively, over $E_0$. After changing of variables, Eqs. (1)-(3) can be presented in the form, where these impedances are unknown variables. Then the results obtained for the gain by direct calculation of the emitted power can be interpreted with the concept of the graphene and superconductor impedances,  characterising the hybrid system. This argument with impedances for graphene and SC has been used just for clarification of the effect and % why we have the gain.  The analysis of impedances 
it provides an alternative interpretation of the gain phenomenon. %, which has been calculated directly from the Eq.(8) for the emitted power, $P ~ < J(r, t)E*(r, t)>$.
\bibitem{remark2} Note, that supporting zero chemical potential we expect to see a nonzero radiation power, which is just an average $V_g(t) \times I(t)$ over photocurrent period.  The output power is maximised when the chemical potential is pinned at the Dirac point, since in this case the voltage variation $V_g(t)$ has a maximal amplitude.  
On the other hand, if we apply an external AC voltage $V_g(t)$ to the hybrid system, the transistor might radiate THz radiation even without external incident light. In this case, the radiation power of the system will be exactly proportional to $P=\langle V_g(t)I(t)\rangle$, where  the average is taken over the oscillation period and here a fraction of this energy will be wasted in thermal Joule losses. 
However, such a source of THz radiation requires a special set up for the applied voltage to maximize the effect of quantum capacitance.
\bibitem{remark-distance} In our calculations we used a relatively large value for the graphene-superconductor separation, $a$=10 nm, which is several orders of magnitude smaller than the wavelength of THz radiation.
The smaller the separation the stronger the effect.  However, at very small separation there may arise a proximity effect. That limits the separation distance to  a superconducting coherence length, which is for HTSC about $\sim$ 1-2 nm.
\end{thebibliography}
\end{document}